\documentclass[12pt]{iopart}

\usepackage{amsfonts}
\usepackage{amssymb}
\usepackage{epsfig}
\usepackage{subfigure}
\usepackage{graphicx}

\def\overleftrightarrow{\stackrel{\leftrightarrow}}

\begin{document}

\title{Photon location in spacetime}
\author{Margaret Hawton}
\address{Department of Physics, Lakehead University,
Thunder Bay, Canada}

\begin{abstract}
The NewtonWigner basis of orthonormal localized states is generalized to
orthonormal and relativistic biorthonormal bases on
an arbitrary hyperplane in spacetime. This covariant formalism 
is applied to the measurement of photon location using a hypothetical
3D array with pixels throughout space turned on at a fixed time and a 
timelike 2D photon counting array detector with good time resolution. A moving
observer will see these detector arrays as rotated in
spacetime but the spacelike and timelike experiments remain distinct.

\end{abstract}
\maketitle

\section{Introduction}

In nonrelativistic quantum mechanics the localized states at all positions in
space at a fixed time form a basis of Dirac delta-functions. In relativistic
quantum mechanics particle localization is a difficult and controversial
concept. Photons are always in the relativisitic regime so the problems are
especially severe in their case. However, in spite of the theoretical
difficulties, localized photon states are a useful concept. The projection
operators onto these localized states define a positive operator valued
measure (POVM) that has been used to describe photon position in imaging and
photon counting experiments \cite{Teich,Tsang,HawtonPOVM}.

While the probability density basis describes the position of a particle at a fixed
time, an experimenter might use a photon counting array detector and record
arrival time at one of its pixels. In the nonrelativistic context a spatial
volume element is a scalar. In relativity space and time form the four
dimensional Minkowski spacetime in which the volume of ordinary space is a
spacelike hyperplane with its normal parallel to the time axis. The elements
of a photon counting array have normals in a spacelike direction and thus are
timelike. Here both possibilities will be combined to give a
covariant theory of photon location.

Newton and Wigner (NW) derived bases of localized states at a fixed time for
massive particles and zero mass particles with spin zero and one-half
\cite{NW}. These states are orthogonal and hence localized in the sense that
the invariant inner product of states centered at different positions equals
zero. The NW procedure failed for photons because spherical symmetry was
assumed. However a basis of localized photon states with axial symmetry can be
constructed using the NW procedure \cite{HawtonPositionOp}, thus extending the
concept of NW localization to photons.

Particle density is usually expressed in terms of the positive frequency part
of the field. According to Hegerfeldt's theorem this leads to instantaneous
spreading and possible causality violations \cite{Hegerfeldt}. Recent work on
the Klein Gordon (KG) equation shows that it is possible to include negative
frequency terms and this approach will be applied to the photon here.

The plan of this paper is as follows: Some recent work on the KG particles
will be reviewed in Section 2 and extended to the photon in 3. Probability
density, photon counting and the perspective of a moving observer will be
examined 4 and we will conclude. Rationalized natural units ($\hbar
=c=\varepsilon_{0}=\mu_{0}=1$) and covariant notation will be used throughout.
An event in spacetime will be described by the four-vector
$x=x^{\mu}=\left(  t,\mathbf{x}\right)  $. With the metric signature $\left(
-,+,+,+\right)  $, $x_{\mu}=\left(
-t,\mathbf{x}\right)  $. The four-wavevector is $k^{\mu}=\left(
k^{0},\mathbf{k}\right)  $. Repeated indices in products of four-vectors will
imply summation and a contraction such as
$kx=k_{\mu}x^{\mu}=-k^{0}t+\mathbf{k\cdot x}$ is an
invariant. Sums over polarization and positive and negative flux directions
will be written out explicitly.

\section{Klein-Gordon particles}

Recent work motivated by attempts to reconcile quantum mechanics with general
relativity has lead to a better understanding of the localization of KG
particles. A norm that is positive definite for both positive and negative
frequencies and zero for their cross terms can be defined as
\cite{HenneauxTeitelboim,HalliwellOrtiz,KoksmaWestra}%
\begin{equation}
\left\langle \mathbf{k},\varepsilon\mathbf{|k}^{\prime},\varepsilon^{\prime
}\right\rangle =2\omega\delta^{3}\left(  \mathbf{k}-\mathbf{k}^{\prime
}\right)  \delta_{\varepsilon,\varepsilon^{\prime}} \label{kNormalization}%
\end{equation}
where $\omega=\sqrt{\mathbf{k\cdot k}+m^{2}}$ and $\varepsilon=\pm$. The KG field can then be
written as%
\begin{equation}
\psi_{\varepsilon}\left(  x\right)  =\int_{k^{0}=\varepsilon\omega}\frac
{d^{3}k}{2\omega}\frac{\exp\left(  ikx\right)  }{\left(  2\pi\right)  ^{3/2}%
}\psi_{\varepsilon}\left(  k\right)  \label{KGWaveFunction}%
\end{equation}
\ where $\left\vert \psi\right\rangle $ is a one-particle state vector,
$\psi_{\varepsilon}\left(  k\right)  =\left\langle k,\varepsilon
|\psi\right\rangle $ and $k^{0}=\varepsilon\omega$. The
invariant inner product defined in \cite{HalliwellOrtiz} is%
\begin{equation}
\left\langle \phi|\psi\right\rangle =i\sum_{\varepsilon=\pm}\varepsilon
\int_{\Sigma}d\sigma^{\mu}\phi_{\varepsilon}^{\ast}\left(  x\right)
\overleftrightarrow{\partial}_{\mu}\psi_{\varepsilon}\left(  x\right)
\label{KGInnerProduct}%
\end{equation}
where%
\begin{equation}
f\left(  x\right)  \stackrel{\leftrightarrow}{\partial}_{\mu}g\left(  x\right)
\equiv f\left(  x\right)  \partial_{\mu}g\left(  x\right)  -g\left(  x\right)
\partial_{\mu}f\left(  x\right)  \label{Derivative}%
\end{equation}
and $\Sigma$ is a hyperplane with normal surface elements $d\sigma^{\mu}$. The
integrand in (\ref{KGInnerProduct}) is the KG particle flux across $\Sigma$.

\section{Photons}

Although (\ref{kNormalization}) is invariant, it will
be generalized to an arbitrary hyperplane to allow consistent evaluation of the 
equations derived here. Also, for photons it is necessary to include polarization $\lambda$. 
For the special case of planes the reciprocal or $k$-space
normals are the same as those in $x$-space so the hyperplanes will still be
referred to as $\Sigma$ in $k$-space. Since $m=0$ for photons the components
of $k$ are related by the dispersion relation $k^{\mu}k_{\mu}=0$ 
in vacuum. The invariant $\int d^{4}k\delta^{2}\left(  k^{\mu}k_{\mu
}\right)  $ can be integrated over the component of $k$ normal to $\Sigma$ to
give $\int_{\Sigma}d\kappa/2\left\vert k_{\Sigma}\right\vert $ where  
$d\kappa$ is a $k$-space hyperplane element,
\begin{equation}
k_{\Sigma}=k^{\mu}n_{\mu}=\varepsilon\left\vert k_{\Sigma}\right\vert
\label{kSigma}
\end{equation}
and $n^{\mu}$ is a unit normal to $\Sigma$. The
$k$-space orthonormality relation then becomes%
\begin{equation}
\left\langle k\mathbf{,}\lambda,\varepsilon\mathbf{|}k^{\prime},\lambda
^{\prime},\varepsilon^{\prime}\right\rangle =2\left\vert k_{\Sigma}\right\vert
\delta^{3}\left(  k-k^{\prime}\right)  _{\Sigma}\delta_{\lambda,\lambda
^{\prime}}\delta_{\varepsilon,\varepsilon^{\prime}}
\label{kPhotonNormalization}%
\end{equation}
where the subscript $\Sigma$ on the $\delta$-function indicates that it is
valid only on the hyperplane $\Sigma$. The one-photon four-potential in vacuum
can be written as%
\begin{equation}
\psi_{\varepsilon}^{\mu}\left(  x\right)  =\sum_{\lambda}\int_{\Sigma,k_{\Sigma}=\varepsilon|k_{\Sigma}|}
\frac{d\kappa}{2\left\vert k_{\Sigma}\right\vert }e_{\lambda}^{\mu}\left(
k\right)  \frac{\exp\left(  ikx\right)  }{\left(  2\pi\right)  ^{3/2}}%
\psi_{\lambda,\varepsilon}\left(  k\right)  \label{PhotonWaveFunction}%
\end{equation}
where $e_{\lambda}^{\mu}\left(  k\right)  $ form a basis of polarization unit
vectors and
\begin{equation}
\psi_{\lambda,\varepsilon}\left(  k\right)  =\left\langle k,\lambda
,\varepsilon|\psi\right\rangle . \label{kAmplitude}
\end{equation}
With the generalized orthonormality condition (\ref{kPhotonNormalization}),
$\varepsilon$ now denotes the direction of photon flux across $\Sigma$. The
components of $k$ on $\Sigma$ take continuous values from $-\infty$ to
$\infty$ and the dispersion relation then requires that the normal component
of $k$ take two values, $\pm\left\vert k_{\Sigma}\right\vert $.

Since the electric and magnetic field operators form the second rank tensor
$F^{\mu\nu}=\partial^{\mu}A^{\nu}-\partial^{\nu}A^{\mu}$, contraction with the
four-potential $A_{\mu}$ in the Lorenz gauge gives the four-vector $J^{\mu
}=F^{\mu\nu}A_{\nu}$. The positive frequency four-flux operator was derived in
\cite{HawtonNumberOp}. Using the Coulomb gauge in vacuum for simplicity \cite{HawtonNumberOp} gives
\begin{equation}
J_{\varepsilon}^{\mu}\left(  x\right)  =\left\langle \phi\left\vert i\hat{\mathbf{A}}^{\dagger}_{\varepsilon}\cdot\overleftrightarrow{\partial}_t\hat{\mathbf{A}}_{\varepsilon}
,-i\hat{\mathbf{A}}^{\dagger}_{\varepsilon}\times\overleftrightarrow{\nabla}
\times\hat{\mathbf{A}}_{\varepsilon}\right\vert\psi\right\rangle. \label{PhotonFlux}
\end{equation}
By inspection it can be seen that the timelike component involves the electric 
field while the spacelike components require the magnetic field. In the latter the 
difference in (\ref{Derivative}) becomes a sum due to the asymmetry of the cross product.
Eq. (\ref{PhotonFlux}) is the four-vector generalization of the KG flux.
The four-potential $\psi_{+}^{\mu}\left(
x\right)  \propto \left\langle 0\left\vert \hat{A}^{\mu\left(  +\right)  }\left(
x\right)  \right\vert \psi\right\rangle $ replaces the positive frequency part
of the invariant KG field.

Since photon and KG flux satisfy the same continuity equation, by analogy
(\ref{KGInnerProduct}) should replaced with%
\begin{equation}
\left\langle \phi|\psi\right\rangle =\sum_{\varepsilon=\pm}%
\varepsilon\int_{\Sigma}d\sigma_{\mu}J_{\varepsilon}^{\mu}\left(  x\right)  .
\label{FluxIntegral}%
\end{equation}
According to the definitions adopted here, when counting photons their
direction of crossing is irrelevant and the factor $\varepsilon$ ensures that
the particle density on $\Sigma$ is positive regardless of this direction. The
sum over $\varepsilon$ in the $J_{\varepsilon}^{0}$ term is a sum over forward
and backward in time but propagation of a photon backward in time can be
reinterpreted as propagation of an antiphoton forward in time. Negative 
frequency photon absorption will be seen as photon emission so that each  
pixel can act as a detector or a source. 

Integration of (\ref{FluxIntegral}) over $d\sigma$ gives
\begin{equation}
\left\langle \phi|\psi\right\rangle =\sum_{\lambda,\varepsilon}\int_{\Sigma,k_{\Sigma}=\varepsilon|k_{\Sigma}|
}\frac{d\kappa}{2\left\vert k_{\Sigma}\right\vert }\phi_{\lambda,\varepsilon
}^{\ast}\left(  k\right)  \psi_{\lambda,\varepsilon}\left(  k\right)  .
\label{PhotonInnerProduct}%
\end{equation}
The inner product can be evaluated in $k$-space or $x$-space as discussed
\cite{HawtonLorentz}, but evaluation in $k$-space using
(\ref{PhotonInnerProduct}) is simpler.

\section{Spacetime location}

Only three of the four components of $k$ can be treated an independent
variables. Usually the spacelike components, $\mathbf{k}$, are taken to be
independent and a localized basis is defined at a fixed time for all points in
space and each polarization $\lambda$. However, a moving observer will not
agree that localization of these states is simultaneous so this basis is not
invariant. Here covariance is achieved by defining a localized basis on an
arbitrary hyperplane.

The generalized NW localized state on $\Sigma$ at $x'$ with polarization 
$\lambda '$ and flux direction $\varepsilon '$ so that $k_{\Sigma}=\varepsilon ' |k_\Sigma|$ is
\begin{equation}
\chi_{x',\lambda',\varepsilon ';\lambda,\varepsilon}\left(  k\right)  =\sqrt{2\left\vert k_{\Sigma
}\right\vert } \frac{\exp\left(  -ikx'\right)}{\left(  2\pi\right) ^{3/2}}
\delta_{\varepsilon,\varepsilon '} \delta_{\lambda,\lambda '}.
\label{LocalizedPhoton}%
\end{equation}
Here the primed indices are fixed while the unprimed indices are summed over in (\ref{PhotonInnerProduct}).
According to the inner product (\ref{PhotonInnerProduct}) these basis states satsify
\begin{equation}
\left\langle \chi_{x'',\lambda '',\varepsilon ''}|\chi_{x^{\prime},\lambda^{\prime
},\varepsilon^{\prime}}\right\rangle =\delta_{\varepsilon '',\varepsilon^{\prime
}}\delta_{\lambda '',\lambda^{\prime}}\delta^{3}\left(  x''-x^{\prime}\right)
_{\Sigma}. \label{Orthogonality}%
\end{equation}
This implies that these states are localized in the sense originally defined
by NW \cite{NW}. The projection of an arbitrary state vector onto 
the localized state (\ref{LocalizedPhoton}) is
\begin{equation}
\left\langle \chi_{x',\lambda ',\varepsilon '}|\psi\right\rangle =\int_{\Sigma,k_{\Sigma}=\varepsilon '|k_\Sigma|
}d\kappa\frac{\exp\left(  ikx'\right)  }{\left(  2\pi\right) ^{3/2}}\frac
{\psi_{\lambda ',\varepsilon '}\left(  k\right)  }{\sqrt{2\left\vert k_{\Sigma
}\right\vert }}. \label{ProjectionOntoLocalizedPhoton}
\end{equation}
The $x$-space completeness relation is
\begin{equation}
\sum_{\lambda,\varepsilon}\int_{\Sigma,k_{\Sigma}=\varepsilon |k_\Sigma|}d\sigma\left\langle \phi|\chi
_{x,\lambda,\varepsilon}\right\rangle \left\langle \chi_{x,\lambda
,\varepsilon}|\psi\right\rangle =\left\langle \phi|\psi\right\rangle
\label{Complete}%
\end{equation}
as can be verified by substitution of (\ref{ProjectionOntoLocalizedPhoton})
and integration over $d\sigma$ to give (\ref{PhotonInnerProduct}). Eq.
(\ref{Complete}) is equivalent to the partition of the identity operator%
\begin{equation}
\hat{1}=\sum_{\lambda,\varepsilon}\int_{\Sigma,k_{\Sigma}=\varepsilon |k_\Sigma|}d\sigma\left\vert
\chi_{x,\lambda,\varepsilon}\right\rangle \left\langle \chi_{x,\lambda
,\varepsilon}\right\vert \label{Identity}
\end{equation}
that defines a projection valued measure (PVM) which is a special case of a POVM.

Substitution of (\ref{LocalizedPhoton}) in (\ref{PhotonWaveFunction}) gives
\begin{equation}
\chi_{x',\lambda',\varepsilon ';\varepsilon}^{\mu}\left(  x\right)  =\int_{\Sigma,k_{\Sigma}=\varepsilon ' |k_\Sigma|}
\frac{d\kappa}{\sqrt{2\left\vert k_{\Sigma}\right\vert} }e_{\lambda '}^{\mu}\left(
k\right)  \frac{\exp \left[ ik (x-x')\right] }{\left(  2\pi\right) ^3}
\delta_{\varepsilon,\varepsilon '}
\label{NotLocalized}%
\end{equation}
demonstrating that the potential describing a particle localized at $x'$ is  
not itself localized as noted by NW. 

Eqs. (\ref{PhotonInnerProduct}) to (\ref{NotLocalized}) are the central results of
this paper. In the following subsections they will be applied to probability
density and photon counting measurements as seen by stationary and moving observers. 
The equations can be generalized to allow counting of $N$ photons \cite{HawtonPOVM}.

\subsection{Probability density}

In quantum field theory (QFT) probability density is usually defined at a
fixed time $t,$ so that the integrals should be evaluated on a spacelike
hyperplane with timelike normal $n^{\mu}=\left(  1,0,0,0\right)  .$ To measure
probability density one can imagine an array of transparent detectors
throughout space turned on at $t=a$ as sketched as in Fig.1. In this case
(\ref{ProjectionOntoLocalizedPhoton}) reduces to%
\begin{equation}
\left\langle \chi_{x,\lambda,\varepsilon}|\psi\right\rangle =\int d^{3}%
k\frac{\exp\left(  ikx\right)  }{\left(  2\pi\right)  ^{3/2}}\frac
{\psi_{\lambda,\varepsilon}\left(  k\right)  }{\sqrt{2\omega}}.
\label{ProbabilityAmplitude}%
\end{equation}
The probability density is $\sum_{\lambda,\varepsilon}\left\langle \psi
|\chi_{x,\lambda,\varepsilon}\right\rangle \left\langle \chi_{x,\lambda
,\varepsilon}|\psi\right\rangle $ and the probability to detect a photon is
its integral over volume. In spacetime this volume is
a hyperpixel with a timelike normal.%

%TCIMACRO{\FRAME{ftbpFU}{3.2673in}{2.6878in}{0pt}{\Qcb{Spacetime diagram
%according to an observer in a reference frame where the experimental apparatii
%are stationary. The horizontal shaded region depicts a spacelike detector array
%turned on at a fixed $t$ while the vertical region depicts a timelike
%photon counting detector at a fixed position, $x_{3}=b.$ The dashed and
%dot-dash lines parallel to the light cone represent the limits of the photon
%probability density. For a spacelike (timelike) measurement this terminates on
%$t=a$ ($x_{3}=b$). The boxes depict the measurement result. }}{}%
%{figure1.eps}{\special{ language "Scientific Word";  type "GRAPHIC";
%width 3.2673in;  height 2.6878in;  depth 0pt;  original-width 10.4063in;
%original-height 8.5513in;  cropleft "0";  croptop "1";  cropright "1";
%cropbottom "0";  filename 'Figure1.eps';file-properties "XNPEU";}}}%
%BeginExpansion
\begin{figure}
[ptb]
\begin{center}
\includegraphics[
height=2.6878in,
width=3.2673in
]%
{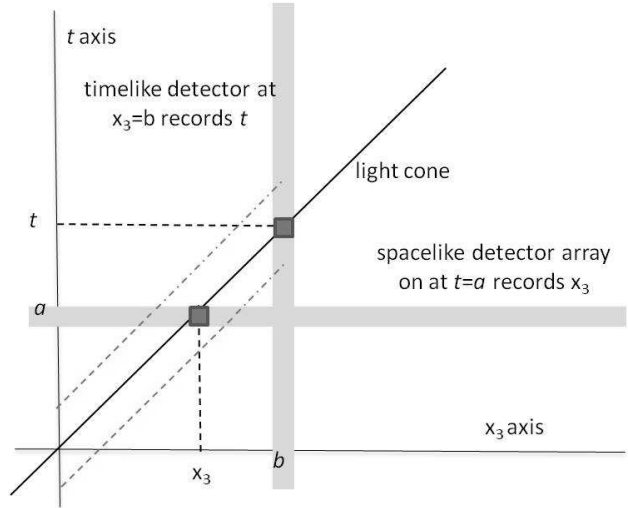}%
\caption{Spacetime diagram according to an observer in a reference frame where
the experimental apparatii are stationary. The horizontal shaded region depicts a
spacelike detector array turned on at a fixed $t$ while the
vertical region depicts a timelike photon counting detector at a fixed position,
$x_{3}=b.$ The dashed and dot-dash lines parallel to the light cone represent
the limits of the photon probability density. For a spacelike (timelike)
measurement this terminates on $t=a$ ($x_{3}=b$). The boxes depict the
measurement result. }%
\end{center}
\end{figure}
%EndExpansion

\subsection{Photon counting \ \ \ \ \ \ }

Since space and time are on an equal footing in a covariant theory, it is
equally valid to fix a spatial coordinate. In fact, this choice describes a
planar photon counting array detector with good time resolution. It will be
assumed that this detector occupies the plane $x_{3}=b$ with normal $n^{\mu
}=\left(  0,0,0,1\right)  $ parallel to the $x_{3}$-axis as also sketched in
Fig. 1. Projection of $\left\vert \psi\right\rangle $ onto this basis gives%
\begin{equation}
\left\langle \chi_{x,\lambda,\varepsilon}|\psi\right\rangle =\int dk_{1}dk_{2}dk^{0}
\frac{\exp\left(  ikx\right)  }{\left(2\pi\right)  ^{3/2}} \frac{\psi
_{\lambda,\varepsilon}\left(  k\right)  }{\sqrt{2\left\vert k_{3}\right\vert
}}.\label{PhotonCounting}%
\end{equation}
The probability to detect a photon is $\sum_{\lambda,\varepsilon}\left\langle
\psi|\chi_{x,\lambda,\varepsilon}\right\rangle \left\langle \chi
_{x,\lambda,\varepsilon}|\psi\right\rangle $ integrated over pixel area and
measurement time, that is integrated over a hyperpixel with a spacelike normal.
If the "wrong" spacelike probability density basis is used to calculate the flux
across a timelike hypersurface an addition factor $cos\theta=k_{3}/ \omega$
arises and the formalism is not covariant \cite{Mostafazadeh2}. The result 
derived here is more like Fleming's covariant generalization of the NW basis \cite{Fleming}.

If an ideal detector is positioned appropriately relative to the source, the
photon will be detected in some hyperpixel of the array with certainty
\cite{HawtonPOVM}. In writing (\ref{PhotonCounting}) it was assumed that the
$k^{0}$ integral extends from $-\infty$ to $\infty$. For an optical pulse
whose line width is small in comparison with its center frequency the negative
frequency contributions are negligible, but negative frequencies can be
included as discussed previously. The wavevectors $k_{1}$ and $k_{2}$ range
from $-\infty$ to $\infty$ but $k_{3}$ takes only the two values $\pm
\sqrt{\omega^{2}-k_{1}^{2}-k_{2}^{2}}$. The wavevector component $k_{3}$ can
be imaginary. For two points in the plane of the detector $k_{3}$ does not
appear in $\left\langle \chi_{x'',\lambda '',\varepsilon ''}|\chi_{x^{\prime},\lambda^{\prime
},\varepsilon^{\prime}}\right\rangle ,$ but the transition amplitude
between arbitrary points does depend on $k_{3}$. The mathematical properties
of this Angular Spectrum Representation that includes both propagating and
evanescent waves have been studied and it have been applied in optics and
acoustics \cite{AngularSpectrumRepresentation}.

\subsection{Moving observer}

In Fig. 2 the observer is travelling with velocity $-\beta$ relative to the
detectors. According to this observer the spacelike photodetector pixels are
not turned on simultaneously and the timelike photon counting array is in motion. 
The detector coordinates are Lorentz transformed to $x'_{3}=\gamma (x_{3}+\beta t )$
and $t'=\gamma (t+\beta x_{3})$ so that $t'=\gamma^{-1}a+\beta x'_{3}$ and
$x'_{3}=\gamma^{-1}b+\beta t'$. In the spacelike experiment the observer sees
its normal rotated to $n^{\mu}=\left(  \gamma,0,0,\gamma\beta\right)  $ while the 
timelike detector is rotated to $n^{\mu}=\left(  \gamma\beta,0,0,\gamma\right)  $. As
$\beta$ approaches unity the light cone $t'=x'_{3}$ at $\alpha=\tan^{-1}\beta=\pi /4$ 
is approached so no observer
sees a timelike experiment as spacelike and vice versa, but any $n$ has a physical
interpretation.

%TCIMACRO{\FRAME{ftbpFU}{3.1004in}{2.4924in}{0pt}{\Qcb{Spacetime diagram as
%seen by an observer with velocity $-\beta$ relative to the detectors.
%According to this observer the spacelike hyperpixels are not turned on
%simultaneously so the detector's world line makes an angle $\alpha=\tan
%^{-1}\beta$ with the $x_{3}$-axis. The timelike detector has velocity $\beta$
%relative to the $x_{3}$-axis so its world line makes an angle $\alpha$ with
%the time axis. }}{}{figure2.eps}{\special{ language "Scientific Word";
%type "GRAPHIC";  maintain-aspect-ratio TRUE;  display "USEDEF";
%valid_file "F";  width 3.1004in;  height 2.4924in;  depth 0pt;
%original-width 10.9745in;  original-height 8.8142in;  cropleft "0";
%filename 'Figure2.eps';file-properties "XNPEU";}}}%
%BeginExpansion
\begin{figure}
[ptb]
\begin{center}
\includegraphics[
height=2.4924in,
width=3.1004in
]%
{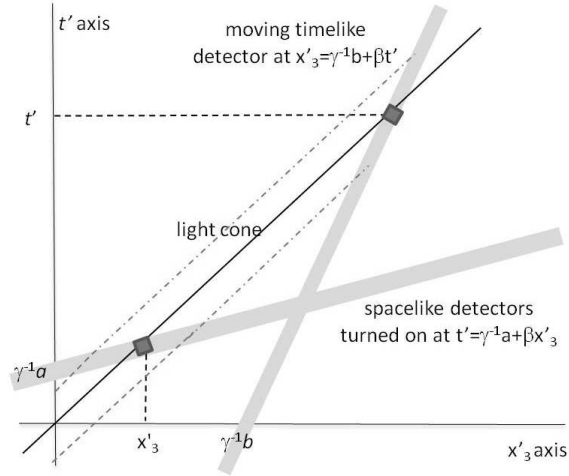}%
\caption{Spacetime diagram as seen by an observer with velocity $-\beta$
relative to the detectors. According to this observer the spacelike
hyperpixels are not turned on simultaneously so the detector's world line
makes an angle $\alpha=\tan^{-1}\beta$ with the $x_{3}$-axis. The timelike
detector has velocity $\beta$ relative to the $x_{3}$-axis so its world line
makes an angle $\alpha$ with the time axis. }%
\end{center}
\end{figure}
%EndExpansion
\ 

\section{Conclusion}

In this paper it was found that it is possible to define an orthonormal NW basis or
a relativistic biorthonormal basis on an arbitrary hyperplane in spacetime.
These bases are localized in the sense defined by Newton and Wigner in 1949 \cite{NW}:
the invariant inner product of two different basis states equals zero. In physical
terms projection of the photon state vector onto one of these bases describes
a spacelike probability density measurement or a timelike photon counting
experiment in which a hyperpixel can act as either a detector or a source. 
Either family of bases is covariant since a moving observer will just
see the hyperplane of the at-rest measurement as rotated in spacetime.
Because the observer is limited by the speed of light, the spacelike
and timelike experiments are distinct. 

Suppose that a source of single photons is available and the location of each
photon is to be measured. The basis of localized states at all points
in spacetime is vastly overcomplete. The importance of hyperplanes is
discussed in \cite{HalliwellOrtiz} and emphasized by Fleming \cite{Fleming}.
Here it was demonstrated that a basis can be defined on any fixed hyperplane
and that such a basis is complete and exclusive. If a photon
exists it should be found somewhere in space at any fixed time (a spacelike
experiment). On the other hand an experimentalist is rather more likely to set
up a photon counting array detector on some plane in space and wait for the
photon to arrive. This is the timelike experiment. In either case it is
logical to look for the photon on a fixed hyperplane. Here it was demonstrated
that the general case incorporating both these possibilities is a covariant
generalization of the NW basis that describes both of these experiments.

\textit{Acknowledgements: }The author thanks the Natural Sciences and
Engineering Research Council for financial support and thanks Juan Leon for
valuable discussions.


\begin{thebibliography}{99} 


\bibitem {Teich}A. F. Avouraddy, B. E. A. Saleh, A. V. Sergienko, and M. C.
Teich, Phys. Rev. Lett. \textbf{87}, 123601 (2001).

\bibitem {Tsang}M. Tsang, Phys. Rev. Lett. \textbf{102}, 253601 (2009).

\bibitem {HawtonPOVM}M. Hawton, Phys. Rev. A \textbf{82}, 012117 (2010).

\bibitem {NW}T. D. Newton and E. P. Wigner, Rev. Mod. Phys. \textbf{21}, 400 (1949).

\bibitem {HawtonPositionOp}M. Hawton, Phys. Rev. A \textbf{59}, 954 (1999); M.
Hawton and W. E. Baylis, Phys. Rev. A \textbf{71}, 033816 (2005).

\bibitem {Hegerfeldt}G.C.Hegerfeldt, Phys. Rev D \textbf{10}, 3320-3321 (1974).

\bibitem {HenneauxTeitelboim}M. Henneaux and C. Teitelboim, Annals Phys.
\textbf{143}, 127 (1982).

\bibitem {HalliwellOrtiz}J. Halliwell and M. Ortiz, Phys. Rev. D \textbf{48},
748 (1993).

\bibitem {KoksmaWestra}J. Koksma and W. Westra, arXiv:1012.3473v1 [hep-th] (2010).

\bibitem {HawtonNumberOp}M. Hawton and T. Melde, Phys. Rev. A \textbf{51},
4186 (1995).

\bibitem {HawtonLorentz}M. Hawton, Phys. Rev. A \textbf{78}, 012111 (2008).

\bibitem {Mostafazadeh}A. Mostafazadeh, J. Mod Phys. A \textbf{21}, 2553 (2006).

\bibitem {HawtonWaveFtn}M. Hawton, Phys. Rev. A \textbf{75}, 062107 (2007).

\bibitem {Mostafazadeh2}A. Mostafazadeh and F. Zamani, Annals Phys. \textbf{321}, 2183 (2006).

\bibitem {Fleming}G. N. Fleming, Philosophy of Science \textbf{67}, 515 (2000).

\bibitem {AngularSpectrumRepresentation}E. Wolf, J. Opt. Soc. Am. A
\textbf{4}, 1920 (1986).


\end{thebibliography}
\end{document}